\documentclass{camera}
\usepackage{graphicx}  

  \renewenvironment{thebibliography}[1]{%
    \begin{oldthebibliography}{#1}%
      \setlength{\parskip}{0ex}%
      \setlength{\itemsep}{0ex}%
  }%
  {%
    \end{oldthebibliography}%
  }

\begin{document}

%
\title{Critical-like behavior in a lattice gas model}

%
\author{A. Wieloch$^{1}$, J. Brzychczyk$^{1}$, J. \L{}ukasik$^{2}$, P. Paw\l{}owski$^{2}$,\\ T. Pietrzak$^{1}$
        \and W. Trautmann$^{3}$}

%
\organization{
$^{1}$ Institute of Physics, Jagiellonian University, 30-059 Krak\'{o}w, Poland\\
$^{2}$ Institute of Nuclear Physics PAN, 31-342 Krak\'{o}w, Poland\\
$^{3}$ GSI, D-64291 Darmstadt, Germany}

\maketitle

\begin{abstract}
ALADIN multifragmentation data show features characteristic of a critical behavior,
which are very well reproduced by a bond percolation model.
This suggests, in the context of the lattice gas model, 
that fragments are formed at nearly normal nuclear densities and temperatures
corresponding to the Kert\'{e}sz line.
Calculations performed with a lattice gas model have shown
that similarly good reproduction of the data can also be achieved at lower densities,
particularly in the liquid-gas coexistence region.
\end{abstract}

%

In nuclear multifragmentation studies the presence of a phase transition is often deduced,
however, its nature is not unambiguously identified.
In small systems the asymptotic behavior is strongly
modified by finite size effects so that
the distinction between first- and second-order phase transitions
becomes very difficult. 
Simulations with one-component lattice gas models have shown that critical-like features
are observed in finite systems not only along the Kert\'{e}sz line
but also inside the liquid-gas coexistence zone, {\it i.e.} the first order phase transition
can mimic critical behavior \cite{Gulm:99:1,Carm:02:1,Gulm:05:1}.

We have performed similar criticality analysis with a more realistic isospin dependent
lattice gas model (LGM) which includes Coulomb interactions.
Criticality is deduced from fluctuations of the largest fragment size.
The fluctuations are measured by cumulants (cumulant ratios) of the probability distribution of the
largest fragment size, such as the skewness, $K_{3}$, and the kurtosis excess, $K_{4}$.
As it was shown in percolation studies, the pseudocritical point in finite systems
is indicated by $K_{3}=0$ and a minimum value of $K_{4}$ \cite {Brzy:06:1}.
Our LGM results are confronted with the ALADIN S114 data
on fragmentation of $^{197}$Au projectiles,
following the procedure used in percolation analysis \cite{Brzy:09:1:AW}.

In LGM a system is composed of $Z_{0}$ protons and $N_{0}$ neutrons
which are distributed on a simple cubic lattice with $L$ sites.
The configuration is described by the site occupation numbers $n_{i}$.
If the $i$-th site is occupied by proton (neutron) then $n_{i}=1$ (-1),
if the site is unoccupied then $n_{i}=0$.
The system density is given as $\rho =  \rho_{0} A_{0}/L$, where $\rho_{0}$ is
the normal nuclear density and $A_{0}=Z_{0}+N_{0}$.
The Hamiltonian has the following form:
\begin{equation}
H=\sum_{i=1}^{L} \frac {{p_{i}}^{2}}{2m}n_{i}^{2}- \sum_{<i,j>} \epsilon_{n_{i}n_{j}} n_{i}
 n_{j}+\sum_{n_{i}=n_{j}=1, \, j\neq i} \frac {I_{c}}{r_{ij}}
\label{eq1}
\end{equation}
where $p_{i}$ is the nucleon momentum, $m$ is the nucleon mass, 
$\epsilon_{n_{i}n_{j}}$ is the nuclear coupling constant
($\epsilon_{11}=\epsilon_{-1-1}=0, \, \epsilon_{1-1}=\epsilon_{-11}=5.33$ MeV),
$I_{c}=1.44$ MeV fm is the Coulomb constant,
and $r_{ij}$ is the distance between sites $i$ and $j$
calculated for the lattice spacing $r_{0}=1.8$ fm \cite{Sama:00:1,Leha:08:1}.
The nuclear interaction acts between nearest neighbors $<i,j>$.
The equilibrium configurations were sampled according to the potential
energy with the Metropolis algorithm for the canonical ensemble.
Given a configuration, the momenta of nucleons were randomly drawn from the Maxwell-Boltzmann
distribution. Clusters were defined using the Pan-Das Gupta prescription:
two neighboring nucleons belong to the same cluster if their relative
kinetic energy is insufficient to overcome the attractive bond \cite{Pan:98:1}.
Free boundary conditions were applied to account for the presence of a surface.
Typically, a sample of $10^4$ events was obtained for each considered system at a given
temperature.

\begin{figure}[ht]
\begin{center}
\includegraphics[scale=0.6]{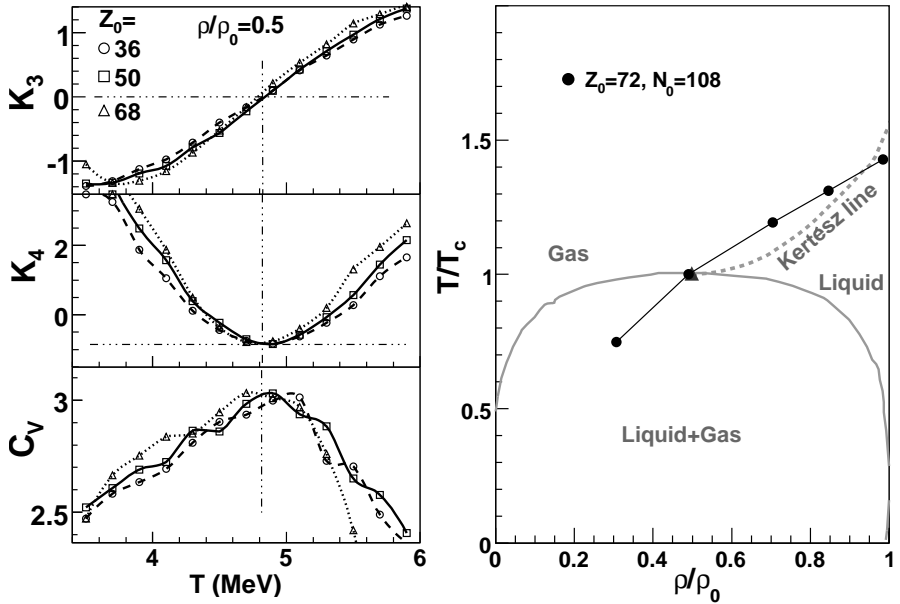}
\caption{Lattice gas model predictions. Left panel: the cumulants of the $Z_{max}$ distribution and the specific heat $C_{V}$
as a function of the temperature at the density $\rho = 0.5 \rho_{0}$ for different system sizes $Z_{0}$.
Right panel: locations of the critical-like points (full circles) in the LGM phase diagram.
The liquid-gas coexistence and the Kert\'{e}sz lines are schematically drawn on the basis
of \cite{Camp:97:1}.}
\label{k3_k4_cv_vs_t} 
\end{center}
\end{figure}

\begin{figure}[ht]
\begin{center}
\includegraphics[scale=0.5]{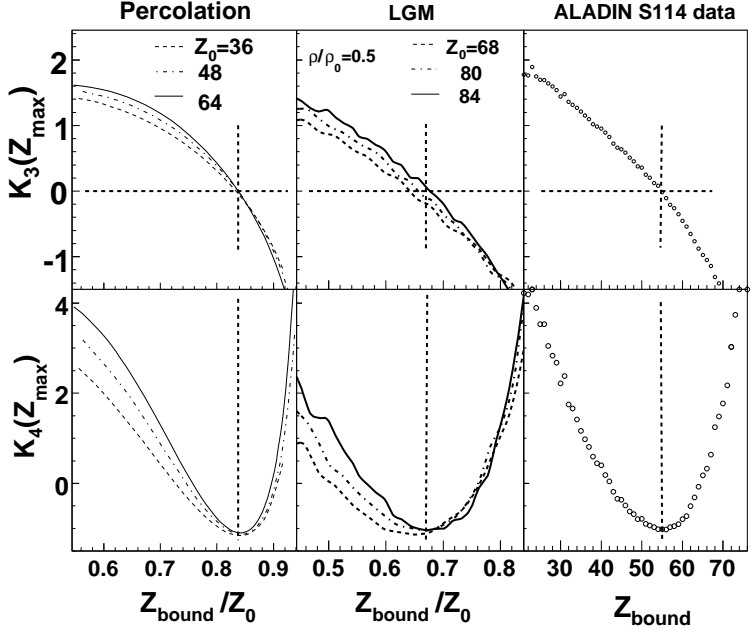}
\caption{The cumulants of the $Z_{max}$ distribution as a function of $Z_{bound}$.
Percolation and LGM calculations are performed for different assumed system sizes $Z_{0}$.
The LGM results are for the density $\rho=0.5 \rho_{0}$.
The experimental data concern the fragmentation of $^{197}$Au projectiles
at relativistic energies.}
\label{k3_k4_vs_zbound} 
\end{center}
\end{figure}

\begin{figure}[h]
\begin{center}
\includegraphics[scale=0.35]{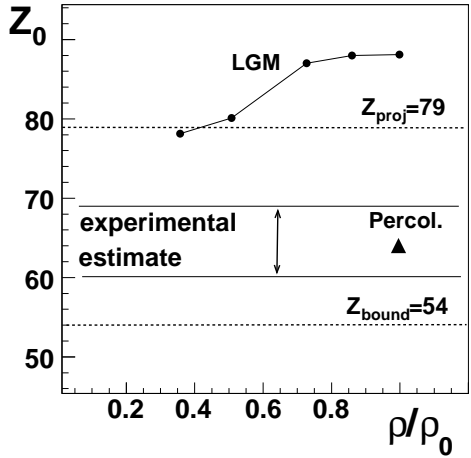}
\caption{The system size $Z_{0}$ at the critical-like point located at $Z_{bound}=54$.
The experimental mean value of $Z_{0}$ is estimated as $64.5 \pm 4.5$ \cite {Poch:95:1}.
The full circles show the LGM results obtained for different densities.
The percolation result is indicated by the triangle.}
\label{zro2} 
\end{center}
\end{figure}

\begin{figure}[h!]
\begin{center}
\includegraphics[scale=0.63]{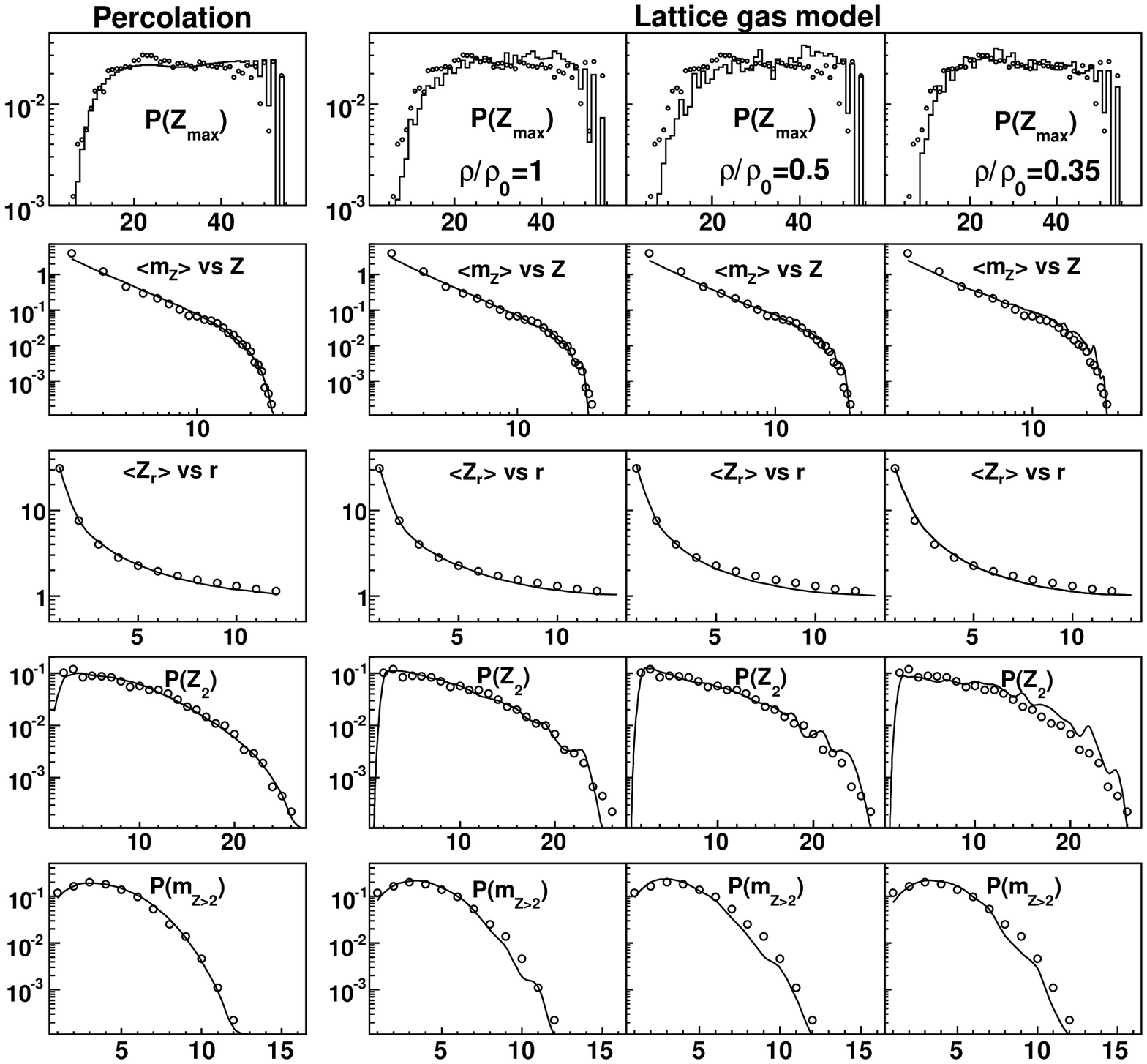}
\caption{Comparisons between the ALADIN data (circles) and models predictions at
$Z_{bound}=54$. The lines represent calculations of the bond
percolation model (first column) and the lattice gas model for densities
indicated in the figure (next three columns). From top to bottom:
the distribution of largest fragment charge, the mean fragment multiplicity
as a function of the fragment charge (largest fragment excluded),
the mean fragment charge as a function of the fragment rank, the probability distribution
of the second largest fragment charge, and the multiplicity distribution
of fragments with $Z>2$.}
\label{lgm_percolation_zdistr} 
\end{center}
\end{figure}

Let us start the examination of the largest fragment size (charge)
distributions in the vicinity of the thermal critical point.
This point is located at the density $\rho = 0.5 \rho_{0}$.
The critical temperature is indicated by a maximum of the
specific heat $C_{V}$. As can be seen from Fig. 1 (the lower left panel)
the critical temperature is about 4.8 MeV for the investigated systems
with $Z_{0}=36 \div 68$. In the present calculations we have assumed the
ratio $A_{0}/Z_{0}=2.5$ of the $^{197}$Au nucleus.
At the critical point, the cumulants $K_{3}$ and $K_{4}$ exhibit
$K_{3}=0$ and minimum of $K_{4}$ as in the case of the percolation transition.
The presence of such criticality features
is also observed for systems at other densities.
We have performed calculations for different density values within the range
$(0.3 \div 1.0) \rho_{0}$. The temperature at the critical-like point
as a function of the system density is shown in the right panel of Fig. 1
for a system with $Z_{0}=72$. This result corroborates that in finite
systems a critical-like behavior may be observed not only along the Kert\'{e}sz line
but also at subcritical densities and temperatures.

In searching for the criticality signal, the control parameter $T$ may
be replaced by the well measured quantity $Z_{bound}$.
The signal ($K_{3}=0$ and minimum $K_{4}$ of about -1)
remains well preserved and appears at a $Z_{bound}$
value corresponding to the critical temperature \cite{Brzy:06:1}.
Figure 2 shows the signal locations on the $Z_{bound}$ axis for percolation,
LGM and the ALADIN S114 data on fragmentation of $^{197}$Au projectile
spectators at the incident energies of $600 \div 1000$ MeV per nucleon \cite{Schu:96:1}.
The model calculations were performed for different system sizes and
the obtained results are plotted against $Z_{bound}$ normalized to the
system charge $Z_{0}$. In such a representation the signal locations
are nearly independent of the system size. In particular, the percolation transition
is observed at $Z_{bound}/Z_{0}=0.84$. 
In the case of the experimental data the size of the fragmenting system is
not precisely established. With decreasing $Z_{bound}$ the average system size systematically
decreases while the excitation energy per nucleon increases. 
As indicated by the cumulants, the criticality conditions are reached
at $Z_{bound}=54$. To observe the criticality signal in percolation at the same
position $Z_{bound}=54$, the system charge must be assumed as $Z_{0} = 54/0.84 \simeq 64$.
In LGM the signal matches the experimental position when
$Z_{0} \simeq 54/0.67 \simeq 80$ in the case $\rho=0.5 \rho_{0}$.
The LGM system charges for this and other densities are shown in Fig. 3.
They are evidently too large, since they exceed the projectile charge.
The percolation result is in a good agreement with an experimental estimate.

In LGM the system sizes are too large with respect to the experimental data,
due to overpredicted multiplicities of $Z=1$ fragments. 
This does not affect $Z_{bound}$ values.
Taking into account only fragments with $Z>1$, the fragment charge partitions
are in agreement with the data, which can be concluded from Fig. 4.
The comparisons shown in Fig. 4 are made at $Z_{bound}=54$ for various observables
related to the fragment charge partitions.
The first column recalls the resemblance between the data and the percolation results \cite{Brzy:09:1:AW}.
In the next columns the same data are compared with the LGM model predictions
for three different densities. In all the cases the reproduction of the experimental
data is astonishingly good.

In summary, the largest fragment charge distributions characteristic of critical behavior,
indicated by $K_{3} = 0$ and minimum $K_{4}$, are observed in LGM
along the Kert\'{e}sz line with an extension to subcritical densities and temperatures.
In the ALADIN data the criticality signal is observed at $Z_{bound}=54$.
The system sizes $Z_{0}$ required by LGM to reproduce the experimental signal are
significantly larger than the experimental estimate.
Considering only fragments with $Z>1$, the simulated characteristics of the fragment charge partitions
at the critical-like point are nearly identical for different freeze-out densities,
and are in a good agreement with the experimental data.
This indicates that the observed differences in $Z_{0}$ values are generated by
different multiplicities of $Z=1$ fragments.
Despite this disagreement, one can conclude that observables related to the fragment charge partitions,
in particular the largest fragment charge distribution, are useful in revealing
a phase transition (critical behavior) in multifragmentation, however,
its location in the phase diagram cannot be unambiguously determined.

Further investigations with LGM are necessary to evaluate how the multiplicity of
$Z=1$ fragments is sensitive to model parameters and cluster definition options.
It will be also interesting to examine isotopic distributions, which might resolve
the freeze-out density question.

\smallskip

This work has been supported by the Polish Ministry of Science and Higher Education grant N202 160 32/4308 (2007-2009).

\end{document}